\documentstyle[aps,prb,epsf,multicol]{revtex}
\begin{document}

%\draft                    % This command makes PACS numbers print

%---------------------------------------------------------------------

\title{
%       \vskip -1.0in\hfill\hfil{\rm\normalsize Printed on \today}
%       \vskip +0.1cm
       First Principles Study of Zn-Sb Thermoelectrics
}

\author{Seong-Gon~Kim}
\address{UES, Inc. and
Code 6691, Complex Systems Theory Branch, Naval Research Laboratory,
Washington, D.C. 20375-5000}

\author{I.I.~Mazin}
\address{CSI, George Mason University and
Code 6691, Complex Systems Theory Branch, Naval Research Laboratory,
Washington, D.C. 20375-5000}

\author{D.J.~Singh}
\address{
Code 6691, Complex Systems Theory Branch, Naval Research Laboratory,\\
Washington, D.C. 20375-5000}

\date{Received Septemper 11, 1997}

\maketitle

%---------------------------------------------------------------------

\begin{abstract}
We report first principles LDA calculations of the electronic structure and
thermoelectric properties of $\beta $-Zn$_{4}$Sb$_{3}$. 
The material is found to be a low carrier density metal with a complex Fermi
surface topology and non-trivial dependence of Hall concentration on band
filling.
The band structure is rather covalent, consistent with experimental
observations of good carrier mobility.
Calculations of the variation with band filling are used to extract the doping
level (band filling) from the experimental Hall number. 
At this band filling, which actually corresponds to 0.1 electrons per 22 atom
unit cell, the calculated thermopower and its temperature dependence are in
good agreement with experiment.
The high Seebeck coefficient in a metallic material is remarkable, and arises
in part from the strong energy dependence of the Fermiology near the
experimental band filling.
Improved thermoelectric performance is predicted for lower doping levels which
corresponds to higher Zn concentrations.
\end{abstract}

\pacs{
73.50.Lw, % Thermoelectric effects
74.25.Fy, % Transport properties (electric and thermal conductivity, 
          % thermoelectric effects, etc.)
71.20.-b, % Electron density of states and band structure of crystalline solids
84.60.Rb  % Thermoelectric, electrogasdynamic and other direct energy conversion
% 72.15.Jf, % Thermoelectric and thermomagnetic effects
% 85.80.Fi, % Thermoelectric devices
% 71.18.+y, % Fermi surface: calculations and measurements; 
          % effective mass, g factor
}

%---------------------------------------------------------------------

\begin{multicols}{2}
%\twocolumn
%\narrowtext

% Introduction

There has been a recent revival of activity in the search for improved
thermoelectric materials, with an emphasis on new materials systems \cite
{Mahan97}.
The efficiency of a thermoelectric is characterized by a dimensionless
figure of merit, $ZT=\sigma S^{2}T/\kappa =S^{2}/L$, where $T$ is the
temperature, $\sigma$ is the electrical conductivity, $S$ is the Seebeck
coefficient (thermopower) and $\kappa$ is the thermal conductivity, which
contains both electronic and lattice contributions, 
$\kappa =\kappa_{\rm el}+\kappa_{\rm lat}$.
The ratio $L=\kappa /\sigma T$, which is often called the Lorentz number, is
ordinarily limited from below by its electronic value, 
$\kappa_{{\rm el}}/\sigma T$, given by the Wiedemann-Franz value, 
$L=(\pi^{2}/3)(k_{B}/e)^{2}$.
Current thermoelectric materials have $ZT\approx 1$.
With the Wiedemann-Franz value for $L$, $ZT>1$ requires $S>160\mu $V/K.
Since doped semiconductors naturally have large thermopower, much of the
thermoelectric research over the past 40 years has focused on covalent
semiconducting compounds and alloys, composed of 4th and 5th row elements, with
a view to finding low thermal conductivity materials that have reasonable
carrier mobilities and high band masses, e.g.  Bi$_{2}$Te$_{3}$,
Si-Ge and PbTe compounds.
Despite the research efforts spanning three decades, little progress in
increasing $ZT$ has been achieved until recently, and in particular
Bi$_{2}$Te$_{3}$/Sb$_{2}$Te$_{3}$ has remained as the material of choice for
room temperature applications.

\begin{figure}[tbp]
\setlength{\columnwidth}{\linewidth} \nopagebreak
\begin{center}
\hbox{
    \epsfxsize=\linewidth
    \epsffile{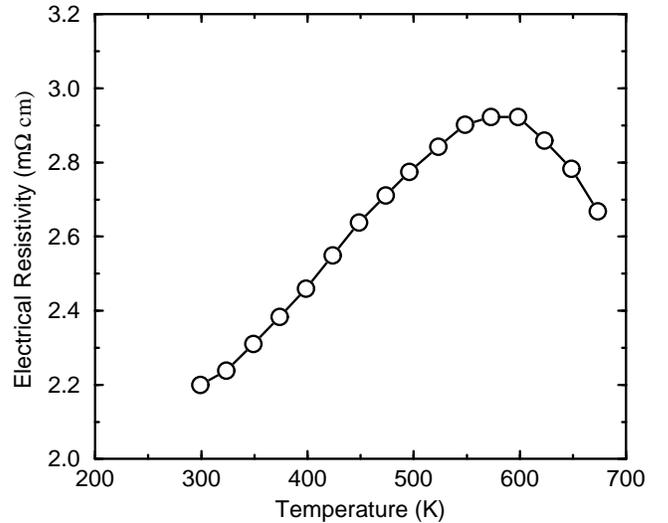}
}
\end{center}
\caption{ Experimental resistivity of $\beta$-Zn$_4$Sb$_3$. 
Replotted with a linear temperature scale 
%using data of Ref.~\cite{Caillat96}. }
using data of Ref.~[2]. }
\label{Fig:resistivity}
\end{figure}

In the last two years, however, three new materials with $ZT \agt 1$ have been
reported \cite {Caillat96,Sales96,Fleurial96}, and these do not clearly fall
into the same class as previous thermoelectric materials.
One of these novel materials, $\beta$-Zn$_{4}$Sb$_{3}$, with reported
$ZT\approx 1.3$ at temperature relevant to waste heat recovery, has a large
region of linear temperature dependence of the resistivity
(Fig.~\ref{Fig:resistivity}), suggestive of a metallic rather than
semiconducting material. But unlike normal metals this is accompanied by high
thermopowers.
% Fig: Resistivity curve.

In this report, we present first principles calculations, within the local
density approximation (LDA), similar to our previous calculations for binary
and filled skutterudites \cite{Singh97,Nordstrom96,Singh94,Feldman96}.
Our main purpose is to understand the rather remarkable thermoelectric
properties of $\beta$-Zn$_4$Sb$_3$ by analysing its transport properties based
on electronic band structures.

% Method

% Fig: Zn4Sb3 crystal structure.
\begin{figure}[t]
\begin{center}
\hbox{
    \epsfxsize=\linewidth
    \epsffile{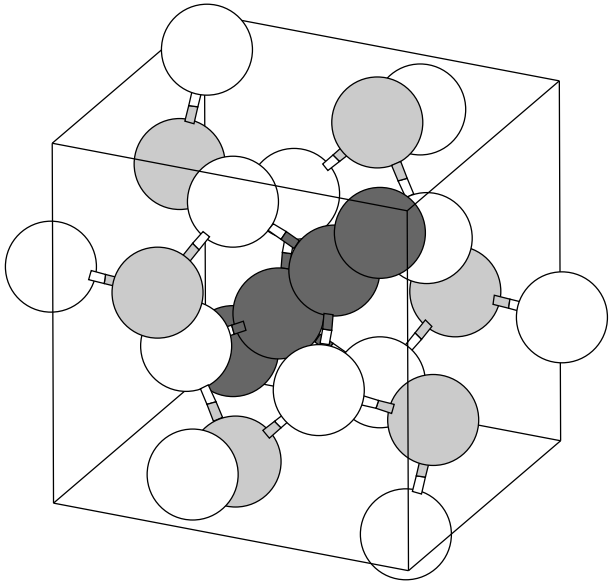}
}
\end{center}
\setlength{\columnwidth}{\linewidth} \nopagebreak
\caption{ Crystal structure of $\beta$-Zn$_4$Sb$_3$.
White and dark grey spheres denotes the 12 Zn and 4 Sb atoms on pure sites in
the 22 atom rhombohedral unit cell. 6 light
grey spheres are occupied by 89\% of Sb and 11\% of Zn atoms. }
\label{Fig:ZnSb}
\end{figure}
\begin{figure}[tbp]
\begin{center}
\hbox{
    \epsfxsize=\linewidth
    \epsffile{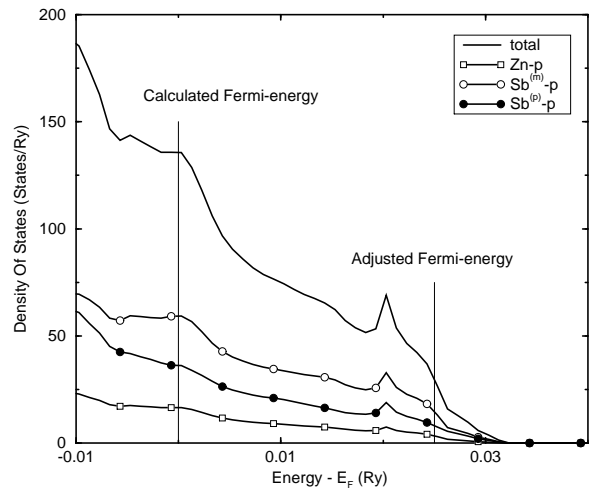}
}
\end{center}
\setlength{\columnwidth}{\linewidth} \nopagebreak
\caption{ Total and partial density of states of $\beta$-Zn$_4$Sb$_3$. 
Total density of states is for 22 atom unit cell and partial density of states
are for each inequivalent atom and magnified by a factor 10 for clarity.
Note the highly covalent character of the band structure.
The calculations were performed for the stoichiometric composition ($E_F$ at
lower vertical line). The actual Fermi energy is given by the upper marker. }
\label{Fig:DOS}
\end{figure}

The computations were performed self-consistently within the LDA using the
general potential linearized augmented planewaves (LAPW) method \cite
{Anderson75,Singh-LAPW}. 
This method makes no shape approximations to either the potential or charge
density and uses flexible basis sets in all regions of space.
As such it is well suited to materials with open crystal structures and low
site symmetries like $\beta$-Zn$_4$Sb$_3$.
Special care has been taken to obtain a well converged basis set of
approximately 2100 functions with LAPW sphere radii of 2.5 a.u. for both Zn and
Sb atoms.
During the self-consistency iterations, the Brillouin-zone integrations
were carried out with 10 special {\bf k} points in the irreducible wedge of
the Brillouin zone; 781 inequivalent {\bf k} points were used to calculate
Fermi surface averages. 
Use of this relatively large number of {\bf k} points was necessary to obtain
accurate Fermi velocities, and, especially, reliable Hall coefficients.
We used the local orbital extension \cite{Singh91} of LAPW method to relax
linearization errors in general and to include the upper core states
consistently with the valence states,
The calculations were based on the experimental crystal structure of
$\beta$-Zn$_{4}$Sb$_{3}$\cite{Pearson} shown in Fig.~\ref{Fig:ZnSb};
however, the site reported to have approximately 11\% of Zn and 89\% of Sb
(light grey spheres in Fig.~\ref{Fig:ZnSb}) in the actual structure
was taken to be a pure Sb site. 
This adjustment yields a formula Zn$_{6}$Sb$_{5}$ with 22 atoms per
rhombohedral unit cell. 
The mixed occupancy was accounted for in a rigid band model as discussed below.
Of the five Sb atoms three occupy the mixed site (Sb$^{(m)}$), and two reside
in a pure Sb site (Sb$^{(p)}$) which is crystallographically inequivalent.
Two Sb$^{(p)}$ atoms form Sb$^{(p)}_{2}$ dimers parallel to the rhombohedral
axis.  The bond length between these two Sb atoms is 2.82 \AA, which is exactly
twice of Sb covalent radius.
This suggest covalent Sb$^{(p)}$-Sb$^{(p)}$ bonding, which is confirmed by the
band structure calculations.
Each Sb$^{(p)}$ forms three additional bonds with Zn, thus being essentially 
four-fold coordinated.
The partial densities of states (Fig.~\ref{Fig:DOS}) shows that
these bonds are also largely covalent.
%
% Fig: Density of states.

Both the Zn and Sb$^{(m)}$ occur in high coordinated
positions; Zn forms one bond with Sb$^{(p)}$, one bond with another Zn, and 
3 bonds with Sb$^{(m)}$, while Sb$^{(m)}$ forms all six bonds with Zn.
The similar coordination of Zn and Sb$^{(m)}$ favors
for Zn to substitute on Sb$^{(m)}$ over Sb$^{(p)}$, as observed.
Substituting one Sb$^{(m)}$ by a Zn
eliminates 6 Zn-Sb bonds and creates 6 Zn-Zn bonds. In general, this is not
an energetically favorable process, and one could expect stoichiometric
compound Zn$_{6}$Sb$_{5}$ rather than Zn-enhanced Zn$_{6.33}$Sb$_{4.77},$ as
observed experimentally. On the other hand, in the stoichiometric compound,
as discussed below, the bonding states of Sb are not yet completely filled,
and doping electrons into the system produces an additional bonding effect.
In other words, substituting Sb by Zn reduces the number of Zn-Sb bonds, but
makes the remaining bonds stronger. 
The balance between the two effects may account for the partial substitution,
however, the very narrow range of compositions for which samples exist remains
unexplained in terms of the calculations.

As mentioned, our calculations were done for the stoichiometric structure. 
The Fermi level lies $\approx 0.4$ eV below the top of the valence band, which
is separated by a sizeable gap (also $\approx 0.4$ eV) from the conduction
bands.
The effect of the deviation from stoichiometry can
be taken into account in a rigid band model: it is assumed that the
electronic structure of the actual compound is the same as for the
stoichiometric one, the only difference being position of the Fermi level.
This is expected to be valid in a highly covalent, broad band system as we
find, and is supported {\it a posteriori} by the comparison with experiment.
Correspondingly we have calculated the relevant transport properties as
functions of the Fermi level position, and used the experimental Hall
conductivity to find the Fermi level position corresponding to actual samples.

The relevant transport properties were determined from the calculated band
structures using the standard kinetic theory as given by Ziman and others
\cite{Ziman72,Hurd72}. For the electrical conductivity, the Bloch-Boltzmann
kinetic equation in lowest order along $x$-direction is (similarly for the
other Cartesian directions) 
\begin{equation}
\sigma_{x}(T) = e^{2} \int d\epsilon \, 
    N(\epsilon) v_{x}^{2}(\epsilon) \tau(\epsilon,T) 
    \left[ -\frac{\partial f(\epsilon)}{\partial \epsilon} \right].
\end{equation}
Here $N(\epsilon)$ is the density of electronic states at the energy 
$\epsilon$ per unit volume, $\tau$ is the scattering rate for electrons,
and the quantity $v_{x}^{2}(\epsilon)$ is defined by: 
\begin{eqnarray}
N(\epsilon) v_{x}^{2}(\epsilon) 
    &=& \frac{2}{(2\pi)^{3}} \int v_{x}^{2} 
            \frac{dS_{\epsilon}}{v_{\epsilon}} \\
N(\epsilon) 
    &=& \frac{2}{(2\pi)^{3}} \int \frac{dS_{\epsilon}}{v_{\epsilon}}
\end{eqnarray}
The integrations are carried out over the iso-energy surface, $dS_{\epsilon}$,
defined by $\epsilon_{\bf k}=\epsilon$. 
For $\epsilon_{\bf k}=E_{F}$, the Fermi energy, the integral is over the Fermi
surface and $v_{\epsilon}$ is the Fermi velocity.
In this case it is related to the square of the plasma frequency
$\omega_{p_{x}}^{2}=4\pi e^{2}N(E_{F})v_{x}^{2}(E_{F})$, which is proportional
to the optical carrier concentration 
$(n/m)_{{\rm eff}}=N(E_{F})v_{x}^{2}(E_{F})$.
We note that in general this does not have a simple relationship to the Hall
concentration or the electron count (i.e. the doping level); 
the Hall concentration is a measure of the average curvature of the Fermi
surface, while the doping level is determined by the volume enclosed by the
Fermi surface.

For isotropic scattering, which is often a reasonable approximation, the
relaxation time does not enter the expression for the Hall concentration,
yielding \cite{Schulz92,Hurd72}: 
\begin{equation}
n_{H}=-\sigma ^{2}/e\sigma _{H}
\end{equation}
with 
\end{multicols}
{\noindent\rule[10pt]{0.5\linewidth}{.1pt}}
\begin{equation}
\sigma_{H} = \frac{e^{3}}{12} \int d\epsilon \, N(\epsilon)
     \mbox{\bf v}(\epsilon) 
     \cdot [{\rm Tr}(\mbox{\bf M}^{-1})-\mbox{\bf M}^{-1}]
     \cdot \mbox{\bf v}(\epsilon) \tau^{2}(\epsilon,T)
     \left[ -\frac{\partial f(\epsilon)}{\partial \epsilon}\right]
     \label{Eqn:Hall}
\end{equation}
\begin{flushright}\rule{0.5\linewidth}{.1pt} \end{flushright}
\begin{multicols}{2}
where for simplicity we have given the expression for cubic symmetry. Here 
{\bf M} is the {\bf k}-dependent effective mass tensor, defined as 
\begin{equation}
\mbox{\bf M}_{\alpha\beta}^{-1} \equiv 
    \hbar^{-1}\frac{\partial v_{\alpha}}{\partial k_{\beta}} \equiv 
    \hbar^{-2}\frac{\partial^{2}\epsilon_{k}}
                   {\partial k_{\alpha}\partial k_{\beta}}
\end{equation}

Provided that the Fermiology does not vary strongly on the scale of 
$k_{\rm B}T$, the derivatives of the Fermi distribution in the above
expressions may be replaced by the $T=0$ limit, which is the delta function,
thereby suppressing the energy integrals.
This approximation was used in the calculations of the Hall concentrations
in this paper.
However, in the expression for the Seebeck coefficient, $S$, one has to include
the energy dependence explicitly at temperatures several times smaller than the
characteristic electronic energy scale.
Thus we have used the full expression,
\begin{equation}
S(T) = \frac{1}{eT\sigma(T)} \int d\epsilon \,\epsilon\sigma(\epsilon,T)
       \left[ -\frac{\partial f(\epsilon)}{\partial \epsilon}\right]
    \label{Eqn:Seebeck}
\end{equation}
where 
\begin{equation}
\sigma(\epsilon,T) = 
    e^2 N(\epsilon) v_{x}^{2}(\epsilon) \tau(\epsilon,T) 
    \label{Eqn:sigma-general}
\end{equation}
is the conductivity corresponding to a Fermi level positioned at $\epsilon$. 
The expression for $\sigma(\epsilon,T)$ in Eq.~({\ref{Eqn:sigma-general}}) 
contains two energy dependent factors; a factor related to the square of
plasma frequency, $\omega_{px}^{2}=4\pi e^{2}Nv_{x}^{2}$ and the relaxation
time, $\tau $. Ordinarily, the first term is the most energy dependent, but
there are exceptions \cite{Karakozov84}, e.g. Pd metal where $E_{F}$ occurs
near a very sharp feature in $N(\epsilon )$ and Kondo systems where there is
resonant scattering \cite{Hurd72}. Since $\beta $-Zn$_{4}$Sb$_{3}$ does not
show any indication of such behavior, we have approximated the energy
dependence of $\sigma $ using only the $Nv_{x}^{2}$ term.

\begin{figure}[tbp]
\begin{center}
\hbox{
    \epsfxsize=\linewidth
    \epsffile{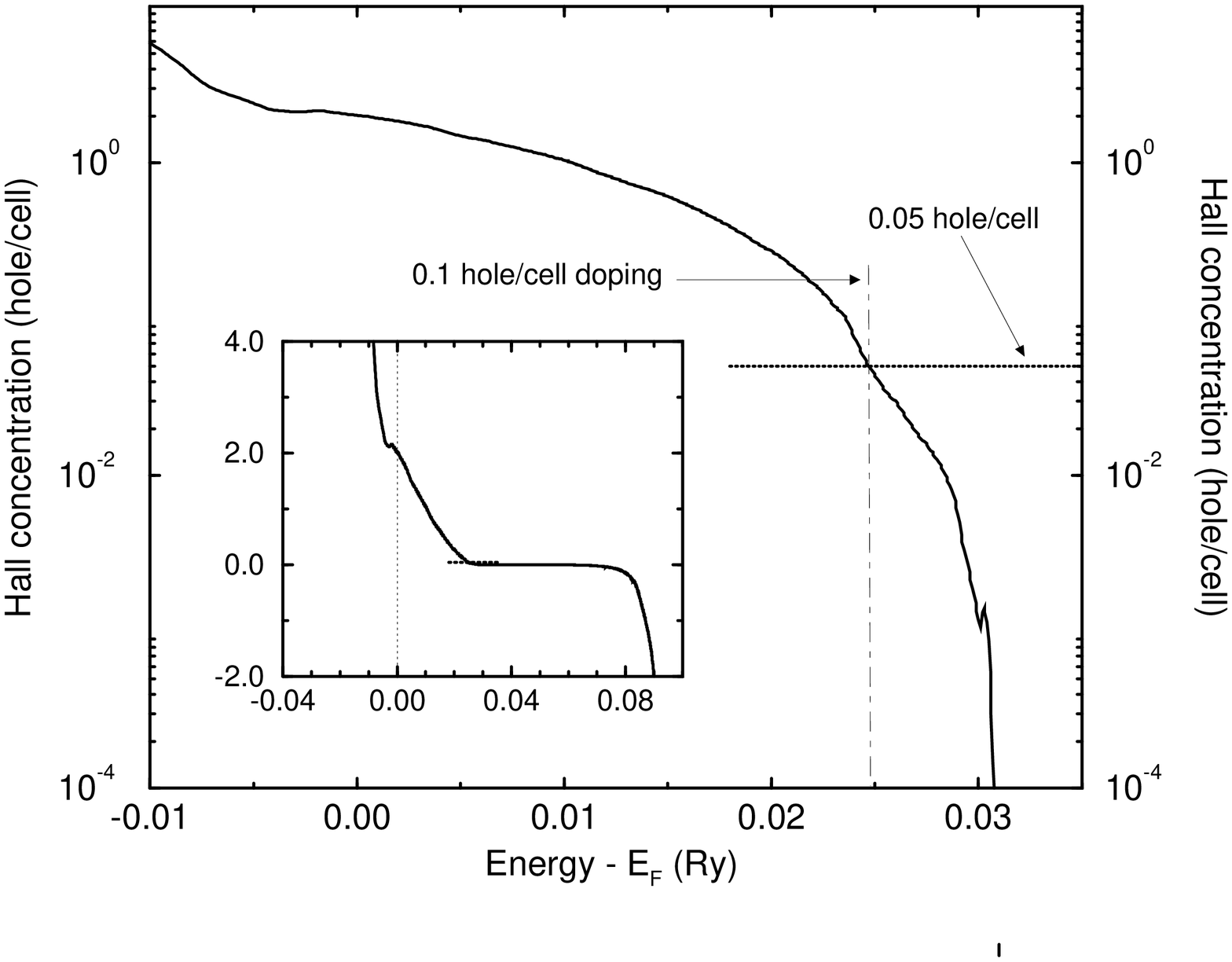}
}
\end{center}
\setlength{\columnwidth}{\linewidth} \nopagebreak
\caption{ Calculated Hall concentration as a function of the Fermi level
position relative to the stoichiometric solid (solid
line) in a linear-log plot. The inset shows the same quantity over wider
range including both sides of the gap. 
The dotted horizontal line represents the experimental value and the
dot-dashed vertical line denotes the required Fermi level shift to match the
calculated Hall concentration to measured value. }
\label{Fig:n_H}
\end{figure}

\begin{figure}[tbp]
\begin{center}
\hbox{
    \epsfxsize=\linewidth
    \epsffile{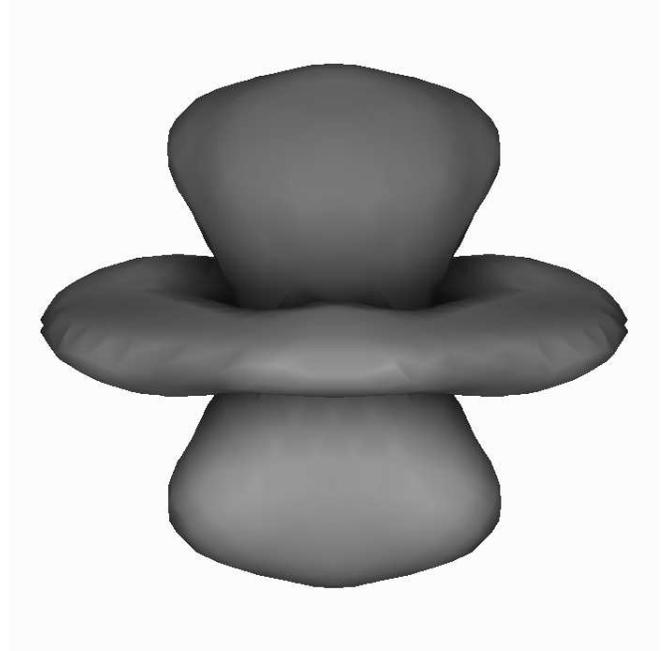}
}
\end{center}
\setlength{\columnwidth}{\linewidth} \nopagebreak
\caption{ Main Fermi surface of $\beta$-Zn$_4$Sb$_3$ at the experimental band
filling. }
\label{Fig:Fermi-surface}
\end{figure}

$\beta$-Zn$_{4}$Sb$_{3}$ has a small Hall number and is characterized either as
a low carrier density metal or a heavily doped semiconductor.
The former terminology is probably more appropriate, because the resistivity
\cite{Caillat96} increases linearly with $T$ over a wide range of at least 
$300\,{\rm K}<T<$550 K (See Fig.~\ref{Fig:resistivity}).
The Hall number, $n_{H}=9\times 10^{19}{\rm cm}^{-3}=0.05$ holes/cell, for the
reported high $ZT$ sample, although small for a metal may be too large to allow
analysis of the transport in terms of usual semiconductor formulae.
The Hall concentration calculated according to Eq.({\ref{Eqn:Hall}}) is plotted
in Fig.\ref{Fig:n_H} as a function of the Fermi level shift from its position
in the stoichiometric compound.
Note that $n_{H}$ is not simply related to the actual number of holes;
the experimental Hall number of 0.05 holes/cell corresponds to the doping level
of 0.1 holes/cell, roughly twice $n_{H}$.
Finally, the calculated plasma frequency at the same position of the Fermi
level is $\omega_{p}=1 \sim 1.2$ eV (depending on polarization), which
corresponds to optical effective carrier concentration $(n/m)_{{\rm
eff}}\approx 0.5(m_{0}/m)$ holes/cell with $m_0$ being the bare electron mass.
Analysis of the Fermi surface (see Fig.~\ref{Fig:Fermi-surface}) shows that
close to the top of the valence band in the range relevant to high $ZT$
samples the main sheet of the Fermi surface is more toroidal rather than
ellipsoidal.
Thus even at this low carrier density,
the Fermi surface has a complicated shape with electron-like and hole-like 
contributions to the Hall coefficient.
This explains why the Hall concentration is so much different from the doping
level, and also suggests that the Fermi velocity and conductivity may depend on
the Fermi level position in a strong and unusual way.
This is the case and it is partially responsible for the high thermopower.
%
% Fig: Hall coefficient.
%
% Fig: Fermi surface
%
% Fig: Thermopower
\begin{figure}[tbp]
\begin{center}
\hbox{
    \epsfxsize=\linewidth
    \epsffile{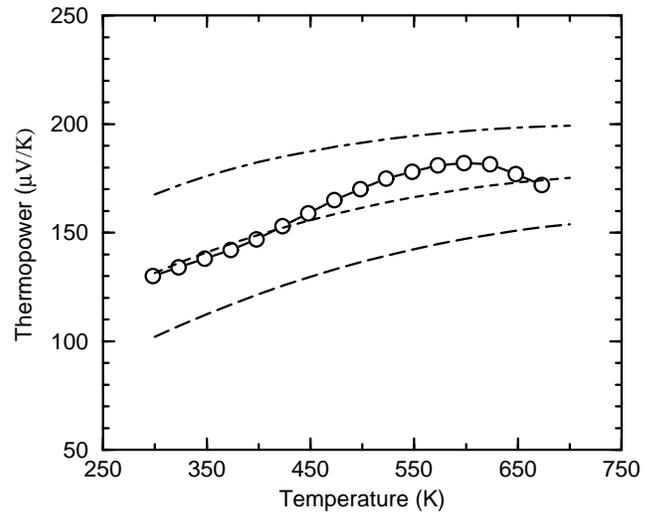}
}
\end{center}
\setlength{\columnwidth}{\linewidth} \nopagebreak
\caption{ Calculated (short dashes) and experimental (solid line with open
circle) thermopower for $\beta$-Zn$_4$Sb$_3$ with 0.05 hole/cell Hall
concentration (0.10 hole/cell doping). $S(T)$ for Hall concentrations 0.12 and
0.02 hole/cell (0.15 and 0.04 hole/cell doping level)
are given by the curves above (dot-dashed) and below (long dashed) of those for
the actual doping level for comparison. }
\label{Fig:thermopower}
\end{figure}

Indeed, using Eq.~(\ref{Eqn:Seebeck}), we find rather high Seebeck coefficient
(Fig.~\ref{Fig:thermopower}).
Naturally, $S$ depends strongly on the position of the Fermi level.
Comparison with the experiment shows that very good agreement is achieved for
the the hole count of 0.1 hole/cell.
This is exactly the hole count that we deduced from the measured Hall number of
0.05 holes/cell, demonstrating the consistency of our approach.

As mentioned, the experimental resistivity up to approximately 550 K
shows typical metallic behavior with high-temperature resistivity
coefficient $d\rho/dT \approx 3.1$ $\mu\Omega\cdot$cm/K and residual
resistivity $\rho_{0} \approx 1.8$ m$\Omega\cdot$cm. 
From these data and the calculated plasma frequency one can estimate the
transport electron-phonon coupling constant $\lambda_{\rm tr}$ and the
scattering rate $\gamma$ due to static defects to obtain
$\lambda_{\rm tr}\approx 1$, and $\gamma \approx 0.2$ eV.
Both numbers are relatively large, but may represent overestimates since they
are not based on single crystal data.
In any case, the experimental data indicate that at high temperature $\kappa$ is
mainly $\kappa_{\rm el}$ in the best reported sample.
This implies that variations in doping level that increase $S$ will also
increase $ZT$. 
Our calculations show not unexpectedly that raising the Fermi energy
corresponding to lower carrier concentration, leads to higher values of $S$,
particularly at intermediate temperatures ranging from room temperature to the
maximum.
This corresponds to higher Zn concentrations on the mixed site.

In summary, we have studied the band structures and transport properties of 
$\beta$-Zn$_{4}$Sb$_{3}$ using first principles electronic structure
calculations. 
According to our {\it ab initio} calculations, the Fermi surface has
complicated topology and this makes the estimation of doping level
non-trivial. 
Combining our calculations with the measured Hall number for the actual
material, we arrive at a relatively large, essentially metallic, value for
carrier concentration, 0.1 hole/cell.
Using this, we calculated Seebeck coefficient and found it to be in excellent
agreement with the experiment both in absolute value and temperature
dependence.
Based upon these results, we characterize this material as metal with complex,
energy-dependent Fermi surface, which provides large thermopower for relatively
high carrier concentration.
This may be useful in identifying other candidate thermoelectric
materials.
The $\beta$-Zn$_{4}$Sb$_{3}$ system is not yet optimized for thermoelectric
application and further increase in $ZT$ is expected if the doping level can be
reduced.
One important issue revolves around the question of why it is so difficult to
make samples with varying Zn concentrations on the mixed site, since higher Zn
concentration would lead to higher $ZT$.
We speculate that the reason has to do with competition from other phases
during the high temperature synthesis.
It would be very interesting to attempt growth by more non-equilibrium
processes such as pulsed laser deposition.

The authors are grateful for helpful discussions with T.~Caillat and 
J.P.~Fleurial.
Computations were performed using DoD HPCMO facilities at NAVO and ASC.
This work is supported by DARPA.
Work at NRL is supported by ONR.

%---------------------------------------------------------------------

\end{multicols}
\end{document}